# Deployment of a Grid-based Medical Imaging Application

S R Amendolia[1], F Estrella[2], C del Frate[3], J Galvez[1], W Hassan[2], T Hauer[1,2], D Manset[1,2], R McClatchey[2], M Odeh[2], D Rogulin[1,2], T Solomonides[2], R Warren[4]

[1]ETT Division, CERN, 1211 Geneva 23, Switzerland
[2]CCCS Research Centre, University of the West of England, Frenchay, Bristol BS16 1QY, UK
[3]Istituto di Radiologia, Università di Udine, Italy
[4]Breast Care Unit, Addenbrooke's Hospital, Cambridge, UK

**Abstract**

The MammoGrid project has deployed its Service-Oriented Architecture (SOA)-based Grid application in a real environment comprising actual participating hospitals. The resultant setup is currently being exploited to conduct rigorous in-house tests in the first phase before handing over the setup to the actual clinicians to get their feedback. This paper elaborates the deployment details and the experiences acquired during this phase of the project. Finally the strategy regarding migration to an upcoming middleware from EGEE project will be described. This paper concludes by highlighting some of the potential areas of future work.

## 1 Introduction

The aim of the MammoGrid project was to deliver a set of evolutionary prototypes to demonstrate that 'mammogram analysts', specialist radiologists working in breast cancer screening, can use a Grid information infrastructure to resolve common image analysis problems. The design philosophy adopted in the MammoGrid project concentrated on the delivery of a set of services that addresses user requirements for distributed and collaborative mammogram analysis. Inside the course of the requirement analysis (see [1] and [2] for details) a hardware/software design study was also undertaken together with a rigorous study of Grid software available from other concurrent projects (now available in [3]). This resulted in the adoption of a lightweight Grid middleware solution (called AliEn (**Ali**ce **En**vironment) [4]) since the first OGSA-compliant Globus-based systems were yet to prove their applicability. Additionally, AliEn has since also been selected as a major component of the gLite middleware for the upcoming EU- funded EGEE (Enabling Grid Environment for E-science) [5] project, as discussed in section 4.

In the deployment phase, the AliEn middleware has been installed and configured on a set of novel 'Gridboxes', secure hardware units which are meant to act as the hospital's single point of entry onto the MammoGrid. These units are configured and tested at all the sites including CERN, Oxford, and the hospitals in Udine (Italy) and Cambridge (UK). While the MammoGrid project has been developing, new layers of Grid functionalities have emerged and hence this has facilitated the incorporation of new stable versions of Grid software (i.e. AliEn) in a manner that catered for a controlled system evolution that provided a rapidly available lightweight but highly functional Grid architecture for MammoGrid.

The MammoGrid project federates multiple (potentially heterogeneous) databases as its data store(s) and uses open source Grid solutions in comparison to the US NDMA [6] project which employs Grid technology on centralized data sets and the UK eDiamond [7] project which uses an IBM-supplied Grid solution to enable applications for image analysis. The approach adopted by the GPCALMA project [8] is similar to MammoGrid in the sense that it also uses AliEn as the Grid middleware but in addition to this it also uses High Energy Physics (HEP) software called PROOF for remote analysis. While GPCALMA is focusing hospitals at the national scale, MammoGrid focuses mammography databases at the international scale. The current status of MammoGrid is that a single 'virtual organization' (VO) AliEn solution has been demonstrated in a prototype called P1 using the so-called MammoGrid Information Infrastructure (MII) and images have been accessed and transferred between hospitals in the UK and Italy. In addition a gLite-based prototype P1.5 has been recently delivered for evaluation for the MammoGrid clinical community and a future prototype, P2, has been targeted to investigate multiple virtual organisations on the MII.

The structure of this paper is as follows. In Section 2 we describe MammoGrid prototypes (i.e. P1 and P1.5) briefly, and section 3 presents details of the deployment environment. Description of the migration strategy towards the new middleware provided by EGEE project is presented in section 4. And, in section 5, conclusions are drawn with emphasis on potential areas for future work.

## 2 MammoGrid Architectural Prototypes

The development of the MammoGrid architecture has been carried out over a number of iterations within the system development life cycle of this project, following an evolutionary prototyping philosophy. The design and implementation was dictated by an approach which concentrated on the application of existing and emerging Grid middleware rather than on one which developed new Grid middleware. Briefly, the work on the development of the system architecture of the MammoGrid project was led from use-case and data model development while evolving the system and system requirements specification.

In particular, the use-case view in Figure 1, being one of the architectural views of the MammoGrid had a significant role in deriving the other architectural views such as logical, implementation, process, and deployment views following the 4+1 views model of the software architecture [9, 10]. In addition to the architecturally significant MammoGrid use-case, the non-functional requirements elicited and specified (e.g. security, SMF standards, performance, etc.) significantly impacted the development of the implementation and deployments views of the MammoGrid discussed in the next sections.

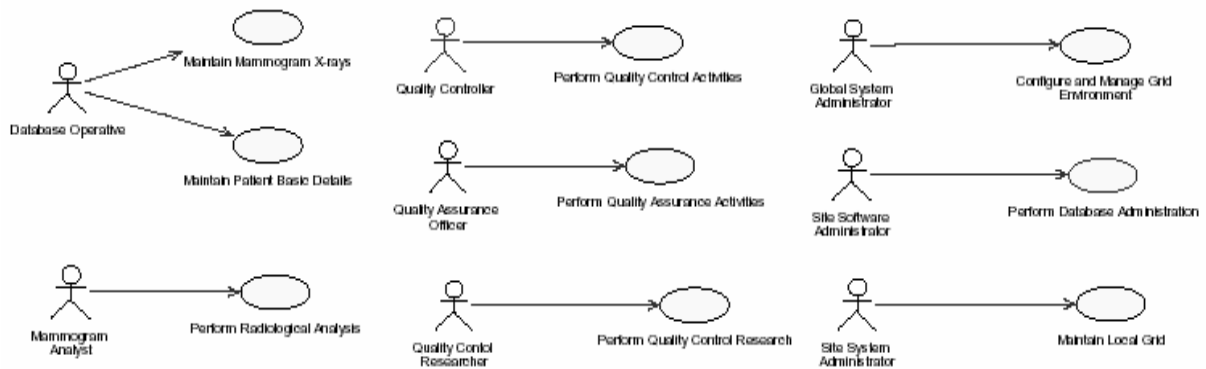

Figure 1: MammoGrid's Main Use-Case View [2]

### 2.1 MammoGrid Architecure: P1 Prototype

The MammoGrid prototype P1 is based on a service-oriented architecture (medical imaging services and Grid-aware services) and has been described in detail in [11] and [12]. It enables mammograms to be saved into files that are distributed across 'Grid-boxes' on which simple clinical queries can be executed. In P1 the mammogram images are transferred to the Grid-Boxes in DICOM [13] format where AliEn services could be invoked to manage the file catalog and to deal with queries. Authenticated MammoGrid clients directly invoke a set of medical image (MI) services and they provide a generic framework for managing image and patient data. The digitized images are imported and stored in DICOM format. The MammoGrid P1 architecture includes a clinician workstation with a DICOM interface to Web Services and the AliEn middleware network. There are two sets of Services; one, Java-based comprising the business logic related to MammoGrid services and the other, Perl-based, the AliEn specific services for Grid middleware. As this architecture is Web Services-based, SOAP messages are exchanged between different layers. RPC calls are exchanged from Java specific services to Alien specific services.

The main characteristics of P1 surround the fact that it is based on a Services Oriented Architecture (SOA), i.e. a collection of co-ordinated services. This prototype architecture is for a single 'virtual organisation' (VO), which is composed of the Cambridge, Udine, and Oxford sites and a central server site at CERN. Each local site has a MammoGrid grid box. As mentioned earlier the Grid hosting environment (GHE) is AliEn, which provides local and central middleware services. The local grid box contains the MammoGrid high level services and local AliEn services. The central grid box contains the AliEn central services. Further details of this prototype are available in [11, 12].

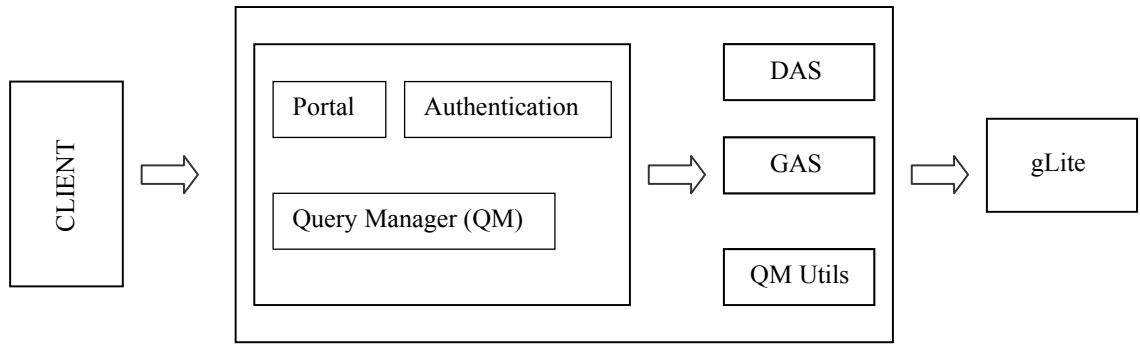

**Figure 2: Prototype P1.5 Components**

## 2.2    MammoGrid Architecure: P1.5 Prototype

The MammoGrid P1.5 architecture, schematically shown in Figure 2, consists of an improved modular design of the MammoGrid software, enhance query handling, coupled with the first release of the upcoming middleware called gLite [5]. In the P1.5 architecture, the databases are distributed and are outside of the gLite software. Consequently the Grid is mainly used to for job execution and to manage files between sites (transfer and replication). There are three different kinds of databases in P1.5 - local database, meta-data database and grid database. The local database stores the local data comprising the local patient's associated data. The meta-data database stores the data model of the local database. The grid database stores gLite/middleware related information.

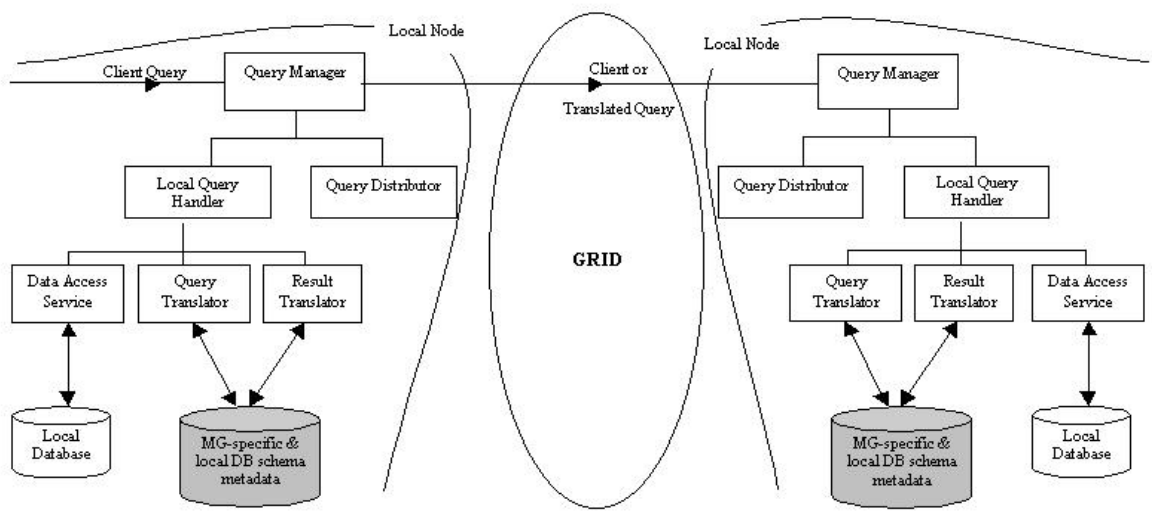

**Figure 3: Implementation View of the P1.5 Prototype: Query Handling Context**

Since the databases and associated metadata are distributed, the queries are executed in a distributed environment. In this context the new services included in P1.5 (see Figure 3) include a Query Manager (QM) service, which is used to manage the queries locally. There is one QM per database per site and it consists of two main components: the Local Query Handler which is responsible for executing the query locally and the Query Distributor that is responsible for distributing the queries to other sites in case of Global queries. Additionally there is a service called QueryTranslator, which translates the client-defined query coming from the above layers (e.g. in XML format) into a SQL statement. Similarly the ResultSet translator translates the results of query execution at local/global site into the desired XML format.

**2.3   Comparison of P1 and P1.5**

Since the Prototype P1 was designed to demonstrate client- and middleware-related functionality, the design of MammoGrid API was kept as simple as possible using a handful of web-service definitions. In essence, the focus of the P1 architecture was centred on the idea of using the existing technologies "as is" and providing the basic functionalities. Since AliEn was the selected Grid middleware, the API design was largely dependent on the features provided by its design. For example, one of the major constraints was regarding the fact that the database was tightly coupled with the file catalogue and the design of AliEn also dictated a centralized database architecture.

The MammoGrid requirement analysis process revealed a need for hospitals to be autonomous and to have 'ownership' of all local medical data. Hence in the P1.5 architecture, the database has been implemented outside of the Grid. The main reason is that having the database environment within the Grid structure made it difficult to federate databases to the local sites using the AliEn software. The only possibility was to change the design of the middleware, which is outside of the scope of the MammoGrid project.

The P1.5 design follows a distributed database architecture where there is no centralization of data. All the metadata are federated and are managed alongside the local medical data records and queries are resolved at the hospitals where the local data are under curation. Another main feature of the P1.5 architecture is the provision of distributed queries. Since the data resides at each hospital site, the queries are executed locally; moreover the clinician can choose to execute a local or global query.

The P1 architecture is rather tightly coupled and it was intended that the P1.5 architecture should be loosely coupled as a good software design principle. Tight coupling is an undesirable feature for Grid-enabled software, which should be flexible and interoperable with other Grid service providers. Thus a key requirement of the P1.5 (and future architectures) is to fully conform to loosely coupled design principles. For example the data layer was loosely separated from the high-level functionality so that the security requirements can be defined and maintained transparently.

In the P1 architecture, an intermediate layer was introduced between the MammoGrid API and the Grid middleware. The purpose of this layer was to provide an interface between the application layer and the Grid middleware layer. Additionally this also served the purpose of providing a gateway to MammoGrid specific calls. This layer was rather complex and its design was not fully optimised such that some of the core MammoGrid API calls were implemented inside its territory. In the P1.5 design these MammoGrid specific calls are migrated from this thick layer to the application layer. This feature was mainly introduced to ease the programming of the MammoGrid API. However this was done at the expense of performance as this increases the SOAP calls from the MammoGrid API to the Grid-aware services.

The above-mentioned factors have had a significant impact on the performance of P1.5 architecture. Since the database is outside the Grid so the overhead of Grid layers is eliminated. Similarly the distributed query model enhances the performance because first the query is executed locally and if desired the query is distributed to the other nodes in the Grid. Although the migration of the MammoGrid specific calls was done at the expense of performance its overall impact is covered by the other factors. Hence it is expected that in terms of performance the P1.5 design will be better than the P1 (see 3.2 section).

# 3 Deployment and operation

## 3.1 Physical Layout

The deployment of P1 is based on a Central Node (CN at CERN) which is used to hold all AliEn metadata, its file catalogue, and a set of Local Nodes (LNs at Oxford, Cambridge and Udine) on which clinicians capture and examine mammograms using the clinicians' workstation interface. The list of services, which should be running on the CN and on the LNs, is shown in Figure 4.

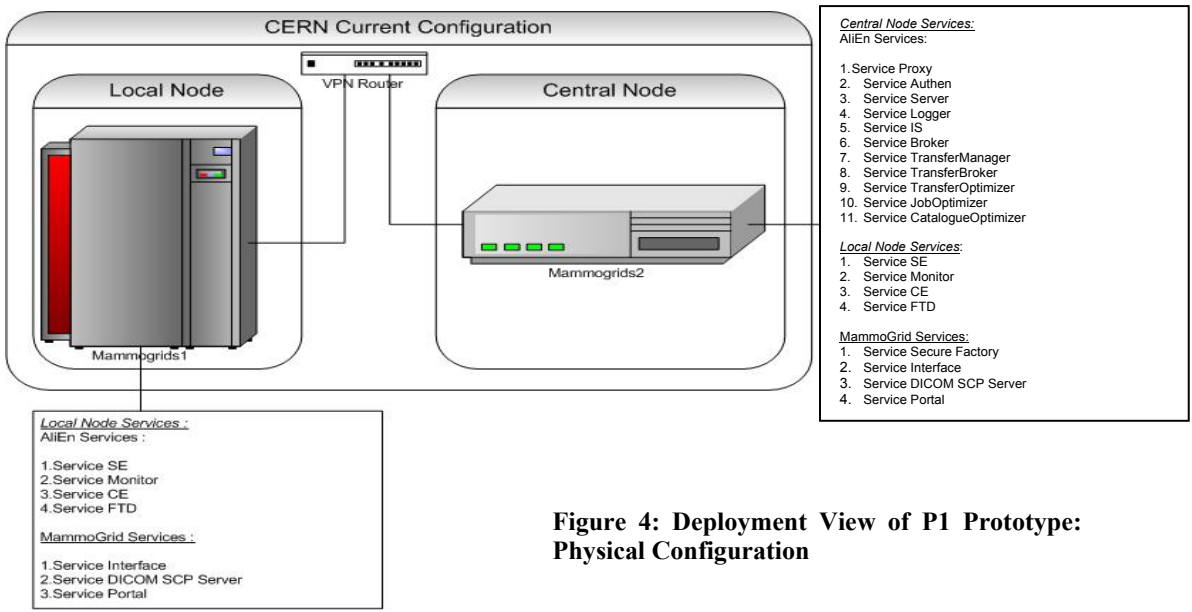

**Figure 4: Deployment View of P1 Prototype: Physical Configuration**

These services include both the AliEn services and the MammoGrid services. The CN and LNs are connected through a Virtual Private Network (VPN). The actual deployment is depicted in Figure 5.

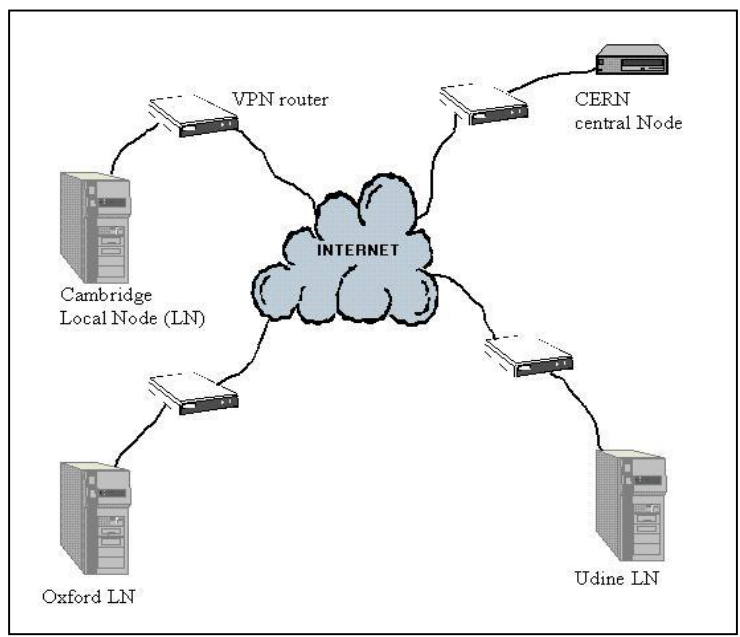

Figure 5: The actual deployment scenario

In order to perform a clean installation on the sites, a test environment has been created at CERN to verify all the installation steps. One important point to be noted was that the Grid Box had to be installed at the actual site from scratch as the IP address of the machine changes from the local site to the actual site. And all the host certificates were needed to be re-issued. The re-issuance could only take place once the IP address has been allotted to the machine on the actual site.

**3.2 Deployment Issues**

As progress has been made in the definition of new Grid standards, AliEn has also undergone evolution in its design. Building a Grids infrastructure in such dynamic circumstances is consequently challenging. In the development and testing phase of MammoGrid it was observed that most of the time bugs were found because of the inconsistencies in the AliEn code. It should be noted that AliEn is made out of several open source components and during the compilation phase it downloads most of the open source components directly from the Internet and this can give rise to incompatibilities and inconsistencies between different components. In order to deal with this situation a design strategy has been adopted. This strategy was different for P1 and P1.5 prototypes. In P1, the working and tested AliEn version has been frozen for our normal development works. In P1.5, in addition to the previous task, the newer AliEn versions were tracked and a move to the newer version was made once this particular version was verified and tested.

As mentioned in section 2, the MammoGrid project has adopted an evolutionary prototyping approach, so by the end of the first year the Prototype P1 was ready and measures were taken for the deployment of this prototype. In the meantime, work on the second prototype had been started. By the time the MammoGrid P1 prototype was deployed in the actual hospital setting the P1.5 architecture was well advanced and was ready to be tested. The initial testing was done in a special testing environment but later it was required to do this in the clinical environment. In order to provide ease of migration a strategy has been adopted in which both the environments could be invoked anytime with minimum effort. According to this strategy the required directories (in the Linux file system), which are used during installation, were mounted and unmounted on the desired locations so as to switch between the two environments.

During the deployment the MammoGrid actual VO has been used for P1. And for P1.5 a Test VO has been created and all the development and testing for the new architecture was done under this Test VO. It should be noted that the deployment of P1 was AliEn based and the current setup of P1.5 is gLite based (see section 4 for details). The Mammogrid project is keeping track of changes in the gLite code such that it is also involved in providing feedback to the developers of the new middleware.

Enabling secure data exchange between hospitals distributed across networks is one of the major concerns of medical applications. Grid addresses security issues by providing a common infrastructure (Grid Security Infrastructure) for secure access and communication between grid-connected sites. This infrastructure includes authentication mechanisms, supporting security across organizational boundaries. One solution regarding the Grid box network security is a hardware-based device, which is, a highly integrated VPN solution. The VPNs could be established both in software and hardware. In the case of software VPNs the entire load of encryption and decryption will be borne by the Grid box, resulting in reduced performance. In the case of hardware VPNs, a specialized device called VPN router is used which is responsible for doing all the encryption, decryption and routing. The VPN router is able to use most connection types, and meet the most demanding security requirements. The measured network throughput when this device is connected to the Grid box with 100 MB WAN port is 11 MB /sec which shows an excellent speed suitable for MammoGrid needs.

In order to preserve privacy, patient's personal data is partially encrypted in the P1 architecture of MammoGrid thereby facilitating anonymization. As a second step in P1.5, when data is transferred from one service to another through the network, communications will be established through a secure protocol HTTPS with encryption at a lower level. Each Grid box will be made a part of a VPN in which allowed participants are identified by a unique host certificate. Furthermore the access from one GridBox to another has been restricted to a given IP addresses. In addition to that only relevant communication ports (AliEn ports and remote admin ports) are opened. This has been done by configuring the router's internal firewall.

Lastly, the P1 architecture is based on trusting the AliEn certification authority. However the P1.5 architecture will improve on these security issues by making use of MammoGrid's own certification authority.

**3.3 Testing procedure**

The deployment of the MammoGrid Information Infrastructure prototypes was a very crucial and important milestone. This phase has so far helped greatly in gaining an insight into the practicality of the medical problem at hand. Some unknown and unresolved issues have been exposed. This section describes the testing procedure followed by some comments on results and performance.

In order to perform tests regarding the deployment, a testing plan has been designed in which all aspects of the testing environment were considered. These tests can be divided into two major parts. The first part deals with the actual testing of the middleware calls. This means that the middleware deployed on all sites has been tested for consistency and performance. These tests included:

a) Replicating files between all the sites (i.e. "mirror" command of the middleware)
b) Retrieving files between sites in both ways (i.e. "get" command of the middleware)
c) Viewing files from a remote site in a local site (i.e. "cat" command of the middleware)
d) Executing jobs on all the nodes (i.e. "submit" command of the middleware)

Similarly other sets of tests included tests of the MammoGrid portal's calls. These tests included loading the DICOM files from the portal, retrieving the files, executing complex queries and executing jobs. The job execution was tested both from the portal and from inside AliEn.

The tests were conducted during the daytime (i.e. the actual operational hours) and by using the standard DICOM files of sizes approximately equal to 8.5 MB. It is obvious that the transfer speed between the nodes plays an important role in terms of overall performance. Furthermore, the Grid overhead was apparent while transferring the data between the nodes as compared to the performance of the direct transfer protocols. This was an expected behavior because the use of Grid obviously places some overhead as different services are called in the complete workflow. This is the price that is required to be paid in order to gain the benefits of Grid such as automatic distribution and scheduling of the jobs, accessing of the resources in various administrative domains and acquiring the collective goal in collaboration.

Job execution on each node was also performed. The jobs, which have been distributed through the Grid, demonstrated the usefulness of Grid, as running the same jobs manually on different nodes was very time consuming. One important point that has been noted during execution of the jobs is that they require to be distributed optimally. For example the jobs were not assigned to the nodes where the actual data was residing instead the selection of nodes was done randomly. This aspect is expected to improve in gLite middleware to achieve optimal performance. It should be noted that the job execution was mainly focused on executing algorithms related to the Standard Mammogram Form (SMF) and Computer Aided Detection (CADe) algorithms of MammoGrid.

In the current deployed architecture the query is centralized because of the centralized database in the P1 prototype as compared to that in P1.5, in which a distributed query execution strategy is undertaken. In the current setup the query handling has been tested for two cases. The first set relates to simple queries and the second includes queries with conjunctions. The simple queries were meant to execute on single tables whereas multiple joins were involved in the conjunctions across multiple tables.

The P1.5 prototype is expected to improve the performance of the overall system especially in all the calls where the database is involved. The Grid related calls will also be improved but we are not expecting any drastic changes. In the next section some of the areas of future work are highlighted.

# 4 Future Works

## 4.1 Migration to EGEE middleware

At the outset of the MammoGrid project there were no demonstrable working alternatives to the AliEn middleware and it was selected to be the basis of a rapidly produced MammoGrid prototype. Since then AliEn has steadily evolved (as have other existing middleware) and the MammoGrid project has had to cope with these rapidly evolving implementations. Additionally AliEn has been considered as the basis of middleware for numerous applications. For example during the early analysis of ARDA (A Realisation of Distributed Analysis) [14] initiative at CERN, AliEn was found to be the most stable and reliable middleware during the production phases of ALICE experiment. In the meantime, work has been started in a new EU-funded project called EGEE, which is developing a new middleware called gLite. The architecture of gLite is based on experiences from several existing middlewares including AliEn. It is expected that in gLite there will be several improvements in the underlying architecture but our recent experience has revealed that the interface of this middleware is similar to AliEn. If this design persists then it will be more or less seamless for projects like MammoGrid to become early adopters of this new middleware. In essence, the selection of AliEn as a core component in EGEE project has put MammoGrid project at an advantage. For example, MammoGrid's team already have expertise to deal with the upcoming middleware i.e. gLite.

## 4.2 gLite - new features

According to the specifications of gLite [15], a unique interface has been provided that will act as a user entry point to a set of core services, which is called the Grid Access Service (GAS). When a user starts a Grid session, he/she establishes a connection with an instance of the GAS created by the GAS factory for the purpose of the session. During the creation of the GAS the user is authenticated and his/her rights for various Grid operations are checked against the Authorization Services.

The GAS model of accessing the Grid is in many ways similar to various Grid Portals but it is meant to be distributed (the Gas factory can start GAS in a service environment close to the user in network terms) and is therefore more dynamic and reflects the role of the user in the system. The GAS offers no presentation layer as it is intended to be used by the application and not by the interactive user. The GAS feature would be very useful as it provides an optimised Grid access that in turn will ameliorate overall performance of the system.

According to [15], in gLite all metadata is application specific and therefore the applications and not the core middleware layer should optimally provide all metadata catalogs. There can be callouts to these catalogs from within the middleware stack through some well-defined interfaces, which the application metadata catalogs can choose to implement. In the version of AliEn, which has been used in the first phase of MammoGrid deployment, the metadata is middleware specific, which not only puts constraints on the design of metadata but also after a series of several AliEn calls, the database is accessed and which in turn puts constraints on the speed of processing.

## 4.3 Prototype 2

The P1 design is based on a single-VO architecture, implying a partial compromise over the confidentiality of patients' data. As mentioned in section 3.2, in the P1 architecture all metadata is centralized in one site and this requires the copying of some summary data from the local hospital to that site in order to satisfy the predicates required for query resolution. Practically this is not feasible other than in specialist research studies where appropriate authority for limited replication of medical data between hospitals can be sought and granted. P1.5 provides an adhoc solution by taking the database out of the Grid to achieve better performance and control of the database. Another aim of P1.5 is to achieve distributed database architecture. The P2 design will be based on a multi-VO architecture where there will be no centralization of data, all metadata will be managed alongside the local medical data records and queries will be resolved at the hospitals where the local data are under governance. This will provide a more realistic clinical solution, which matches more closely the legal constraints imposed by curation of patient data. With the incorporation of a multi-VO set up, the overall metadata will become truly federated inside the Grid and will provide a more secure solution for the medical community.

Furthermore, full adaptation of the Grid philosophy suggests that the federation should not be confined to the database level but should be realized in all aspects of Grid services (including areas such as authentication, authorization, file replication, etc.). The solution, which will be adopted in P2 design, is a federation of VOs (possibly hierarchical in structure), where the boundaries of the organizations naturally define the access rights and protocols for data-flow and service access between 'islands' that provide and consume medical information.

As the P1 and P1.5 designs were based on simple Web Services, interoperability with other middleware was not practically possible. This is mainly due to the fact that different Grid hosting environments require different underpinning technologies; the lack of common communication protocols results in incompatibilities. While OGSA [16] has adopted a service-oriented approach to defining the Grid architecture, it says nothing about the technologies used to implement the required services and their specific characteristics. That is the task of WSRF [17], which is defining the implementation details of OGSA. WSRF is the standard that is replacing OGSI [18], which was the previous standard for specifying OGSA implementation. The WSRF working group had opted to build the Grid infrastructure on top of Web services standards, hence leveraging the substantial effort, in terms of tools and support that industry has been putting into the field in the context of Web Services standard.

The focus of the next MammoGrid milestone will be the demonstration of Grid services functionality and to that end it is planned to implement an (OGSA-compliant) Grid-services based infrastructure. As a result the focus will be on interoperability with other OGSA-compliant Grid services. It is observed that the major trend in Grid computing today is moving closer to true web-services (from OGSI) and we expect a convergence between different Grid protocols in time for the completion of existing MammoGrid web-services design.

## 5  Conclusions

The MammoGrid project is funded by the EU 5th Framework Programme, and is an example of an emerging specialised sub-category of e-Science projects—e-Health projects. One of the secondary aims of the MammoGrid project is to deliver generic solutions to the problems faced during development and deployment, both technical and social. This paper is presented in that spirit—in addition we feel that further issues highlighted regarding VO Management for the MammoGrid project will be of interest not only to other mammography-related projects, but also to a wide variety of applications in which Grid middleware is utilized.

The MammoGrid project has deployed its first prototype and has performed the first phase of in-house tests, in which a representative set of mammograms have being tested between sites in the UK, Switzerland and Italy. In the next phase of testing, clinicians will be closely involved to perform tests and their feedback will be reflectively utilised in improving the applicability and performance of the system. In its first two years, the MammoGrid project has faced interesting challenges originating from the interplay between medical and computer sciences, and has witnessed the excitement of the user community whose expectations from a new paradigm are understandably high. As the MammoGrid project moves into its final implementation and testing phase, further challenges are anticipated. In conclusion, this paper has outlined the MammoGrid application's deployment strategy and experiences. Also, it outlines the strategy being adopted for migration to the new middleware called gLite.

## 6  Acknowledgements

The authors thank the European Commission and their institutes for support and acknowledge the contribution of the following MammoGrid collaboration members: Predrag Buncic, Pablo Saiz (CERN/AliEn), Martin Cordell, Tom Reading and Ralph Highnam (Mirada) Piernicola Oliva (Univ of Sassari) and Alexandra Rettico (Univ of Pisa) . The assistance from the MammoGrid clinical community is warmly acknowledged especially that from Iqbal Warsi of Addensbrooke Hospital, Cambridge, UK and Prof. Massimo Bazzocchi of the Istituto di Radiologia at the Università di Udine, Italy. In addition Dr Ian Willers of CERN is thanked for his assistance in the compilation of this paper.